# Pitcher Plant Inspired Biomimetic Liquid Infused Slippery Surface Using Taro Leaf

Rahul Sharma,[a,†] Sankha Shuvra Das,[a,†] Udita Uday Ghosh,[b] Sunando DasGupta,[b] and Suman Chakraborty[a]*,

[a]Department of Mechanical Engineering, Indian Institute of Technology Kharagpur, Kharagpur-721302, India.
[b]Department of Chemical Engineering, Indian Institute of Technology Kharagpur, Kharagpur-721302, India.
[†]Authors have equally contributed to this work.
*email: suman@mech.iitkgp.ernet.in

## Abstract

Bio-inspired anti-wetting surfaces, such as lotus leaf or pitcher plant, have led to the development of stable liquid infused slippery surfaces for various scientific applications. The present work demonstrates the use of biomimetic superhydrophobic surface (inspired from taro leaf) for fabrication of a stable air-liquid film. The taro leaf replica is fabricated on PDMS using two step soft-molding technique, where microstructures of the replicated surface are used to retain silicone oil layer to form a stable slippery surface. The fabricated surface exhibits a low contact angle hysteresis (CAH) of ~2°, with sliding angle (SA) of ~ 1.1°, which further affirms the super-slippery nature of the surface. Furthermore, it also shows excellent self-repairing ability, thermal stability and long-term durability of the oil coating against high shear rates, high impact droplets etc. We thus envisage that the present method of fabrication of slippery surface is simple and economical, and thus useful for various applications such as water drag reduction, self-cleaning, anti-fogging, anti-fouling etc.

## 1. Introduction

In the past few decades, with the increasing interest in the domain of micro/nano-fluidics, bioinspired slippery surfaces have received significant importance both in engineering as well as biomedical field[1–5]. In many studies, liquid-repellent superhydrophobic surfaces have been developed by bio-mimicking different plant surfaces found in nature. Leaves of Lotus (*Nelumbo nucifera*), taro (*Colocasia esculenta*), rice (*Oryza sativa*) and rose (*Rosa Damascena*) petals are some of the typical examples of bio-templates used therein.[6–11]. Primarily, liquid-repellent surfaces inspired by the morphological characteristics of lotus leaves[6,7,12] have been documented extensively. It exhibits low affinity (adhesion) for water and this phenomenon of water repellence[13,14] has been aptly referred to as the "lotus effect". On a physical basis, this occurs due to the entrapment of air molecules between liquid and the solid surface which reduces the interfacial liquid-solid contact area. This in turn translates into contact angles (θ) >150° with favourably low sliding angles <10° which makes these special surfaces ideal for applications that demand properties like self-cleaning[15,16], liquid-repellent[16], icophobic[17,18], anti-fogging[19,20], heat transfer enhancement[19–21], low adhesion or drag reduction and energy generation in microchannels[22–26]. However, commercial deployment of such surfaces is still limited since they are ineffective against low surface tension liquids[27] and extreme ambient conditions like high humidity[28], low temperature (<0°C for water)[29], high pressure[30,31]. Also, these surfaces have questionable performance wherein droplet impact is involved [32,33] as their anti-wetting properties are incapable of sustaining for longer duration. Such failures have been attributed to the droplet pinning through the micro/nano-textured surface which causes the transition of the liquid from the non-wetting stable Cassie-Baxter[34] state to the wetting unstable Wenzel[35] state.

Recently, Wong et al. reported an innovative surface inspired by the structure of the *Nepenthes* pitcher plant which resolved this issue. This new class of surface is the slippery liquid-infused porous surface (SLIPS)[1] that are found to be extremely effective in anti-wetting applications. SLIPS work on the basic premise of replacing the air present in the micro/nanostructured porous layer of superhydrophobic surfaces with lubricating fluids, for example- silicone oil and perfluorinated polyether (PFPE). This thin film of lubricating oil present on the top of the micro/nanostructured layer provides the exceptional anti-wetting properties of the SLIP surfaces[36,37]. This also enables the anti-wetting properties of SLIPS to be more stable under high humidity, and pressure conditions (up to about 680 atm.)[1], low adhesion for ice[1,38–40] compared to their micro/nanostructured non-wetting superhydrophobic counterpart. Apart from this, these surfaces have shown high resistance against bacterial bio-fouling[41,42] and protein adsorption[43] which makes them a promising platform for robust underwater applications. SLIPS were also demonstrated to retain their properties even in case of physical damage and are known to possess the property of self-repairing[1].

Artificial non-wetting slippery surfaces have been fabricated by multiple approaches that can be classified as (a) creation of rough micro/nanostructured patterns on hydrophobic substrates or (b) by chemical modification of hydrophobic substrates with a low surface energy material[44–48] and finally infusing it with the lubricant oil. Till date, various methods such as, sol-gel[48] approach, electro-chemical deposition[38], lithography[1,49], mechanical operations[50], chemical vapour deposition (CVD)[51], phase separation[52], wet chemical reaction[53], electrospinning[54,55] and crystallization control[56] have been used to fabricate slippery surfaces using variedly different substrate materials including hydrophobic polymers[45], metallic compounds[38], inorganic compounds[57], silane-based compounds[58–60].

However, the usage of a bio-template for fabrication of SLIPS will not only be an eco-friendly and sustainable alternative, it is in fact the need of the hour. Majority of the bio-inspired superhydrophobic surfaces fabricated till date rely on the replication from either lotus leaf or rose petal. In this article, we provide an alternative template wherein we demonstrate the fabrication of a SLIP surfaces from the rough textured base obtained by mimicking taro leaf. To achieve this, we propose a two-step, soft moulding technique that has been used to create positive replica of the taro leaf on a polymer surface which was used finally to fabricate SLIPS by infusing lubricant fluids. Further, we also characterize the fabricated SLIP and superhydrophobic surfaces which exhibit static contact angle (SCA) of ~105° and ~155° respectively. Furthermore, the liquid infuser is silicone oil in the present study in contrast to commonly used perfluoropolyethers (PFPE) lubricant oils. This curbs the cost of production and at the same time provides increases the durability of the surface towards icing and fouling[61]. The slippery surfaces fabricated by this method are shown to acquire properties like liquid-repellent, self-cleaning and self-repairing among others.

## 2. Experimental materials and methods

In general, a slippery surface consists of two layers, a textured surface with a multiscale roughness as the base layer and a slippery layer as the top layer. The first stage of the fabrication comprises of fabrication of the microscale rough polymer surface using a highly adhesive taro plant leaf replica, while second stage involves development of a slippery layer over the textured polymer layer fabricated in the first stage.

**Stage I: Fabrication of a biomimetic superhydrophobic surface (taro leaf replica)**

A very simple, facile, two-step, and low-cost soft molding process was followed to fabricate a polymer based synthetic superhydrophobic surface with hierarchical microstructure like that of taro plant leaf. Fresh taro (*Colocasia esculenta*) leaves were obtained from the Horticulture section, Indian Institute of Technology, Kharagpur. Sylgard-184, a two component, PDMS based silicone elastomer (Dow Corning) was used to replicate surface microstructures of the freshly obtained taro leaves. The prepolymer was prepared by mixing silicone elastomeric base and curing agent in 10:1 (w/w) ratio. Prepolymer mixture was then degassed in a desiccator, for 30 minutes, to remove air trapped during mixing process. Subsequently, the degassed mixture was poured on a fresh and clean taro leaf to replicate the surface microstructure on prepolymer and left for ~48 hours for curing at room temperature.

Thereafter, the leaf was carefully peeled off and thus the polymer surface was imprinted with the leaf microstructures, hence forth referred to as the negative replica. The negative replica comprises of micropores (see Fig.1) as the reverse pattern of the original pattern of micropillar structures found on taro leaves. The replication procedure is again repeated with the negative replica, serving as the primary mold in the second step of the soft molding procedure. However, as both positive and negative replicas consist of same polymer material, it becomes very important to prevent cohesive bonding between the replicas during final replication process. To ensure this, a superficial and microscopically thin layer was created on the negative replica by following an intermediate process. The negative replica is subjected to oxygen plasma treatment for about 40s which introduces polar functional groups, mainly the silanol group (Si-OH). Followed by immersion of the negative replica in the sigmacote (undiluted) solution which reacts with silanol groups (Si-OH) on the surface to form a covalent, neutral and hydrophobic film on the top of PDMS surface. This superficial film prevents cohesive bonding between PDMS positive and negative replicas during the second replication. The treated negative replica was allowed to air dry in a chemical hood for ~2 hours. Immediately after this, a freshly prepared degassed pre-polymer mixture was gently poured on to the negative replica followed by degassing, to remove the entrapped air. The polymer layers are then allowed to cure in a hot air oven at 95° for 2 hours followed by cooling until it reaches to room temperature. Afterwards, the negative replica is gently peeled off from the sample to obtain a positive replica of the taro leaf surface, containing similar microstructure (micro-pillars) to that of the taro leaf.

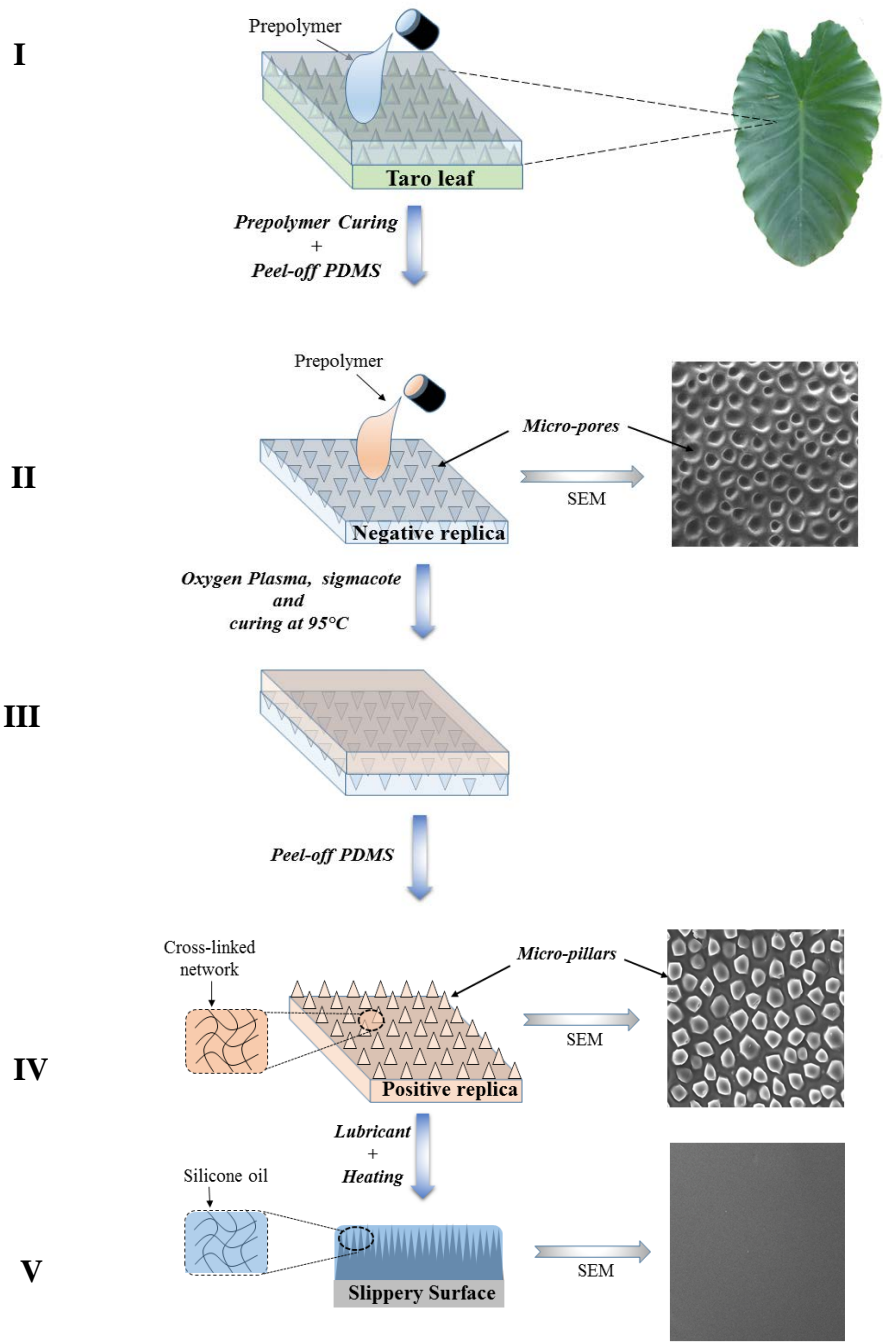

Fig. 1 Schematic representation showing the steps followed during fabrication of SLIPS from superhydrophobic positive replica obtained by replicating surface microstructure of taro plant leaf. (I– fresh and clean taro leaf plant, II– Negative replica of taro leaf, III– Negative and positive replicas during curing process, IV– Positive replica having micropillar structure similar to that taro plant leaf, V– Slippery surface is obtained after infusing positive replica with lubricant oil).

Steady contact angle (SCA) of the sessile droplet for the synthetic surface (positive replica) is obtained as $150° \leq \theta_{SCA} \leq 162°$, which clearly shows the superhydrophobic nature of the replica (see Fig. 2a) (measured by sessile droplet method using a goniometer, Data-physics Instruments GmbH, Germany). The SCA for the corresponding slippery surfaces reduces to a very low value (~105°) but the extremely low CAH (~2.0°) and SA (~1.1°) values aid its super slippery nature which is essential for the droplet mobility (see Fig. 2b). The surface exhibited extremely high contact angles even when the surface was tilted to 180° but the droplet remains suspended at the surface. This indicates high adhesion caused by pinning of the liquid droplets on the textured sites for the replicated surface (see Fig. 2c). The details of the microstructures present on the positive replica was examined and characterized by using a field emission scanning electron microscope (see Fig. 3).

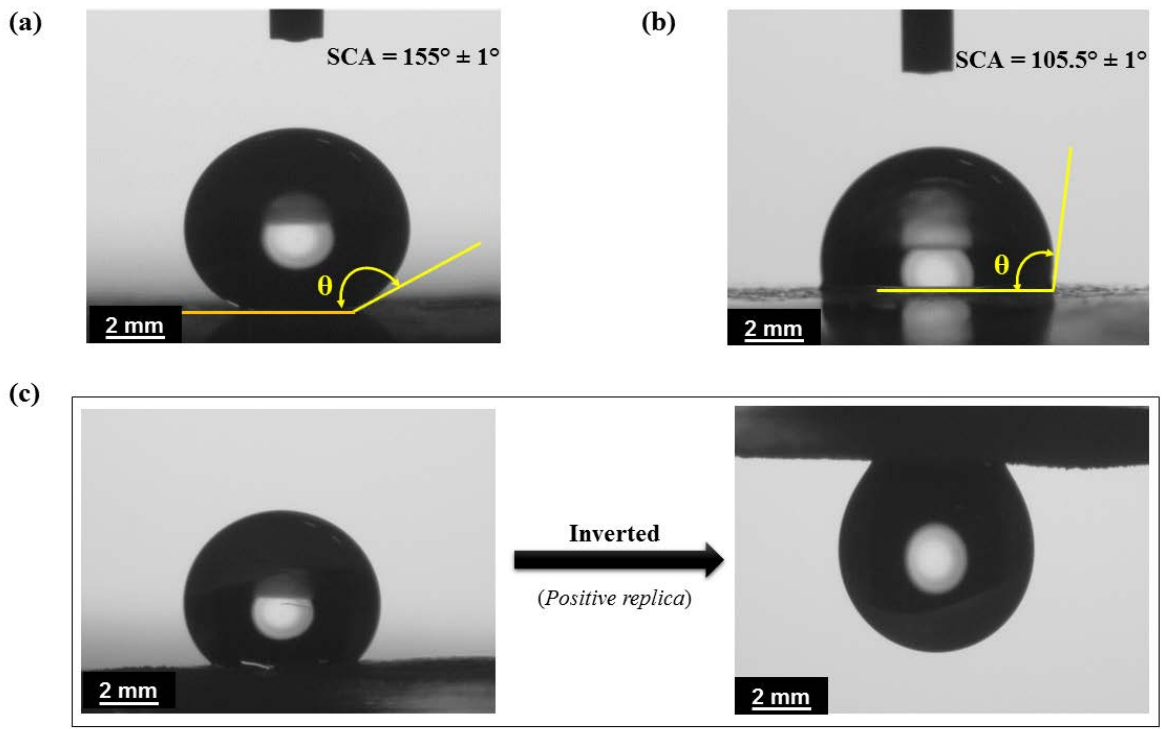

Fig. 1 (a) Positive replica of taro plant leaf with SCA ~155° shows its superhydrophobic nature, (b) SLIPS although have low SCA (~ 105°) but CAH and SA values reaches to remarkably very low values (CAH ~ 2.0° and SA ~ 1.1°), which confirms their super-slippery nature, (c) Synthetic surface (*positive replica*) retains water droplets even when inverted, which indicates high adhesion at the surface and high contact angle hysteresis (CAH) caused by pinning of the liquid droplets on the microtextured sites of the surface.

To reduce adhesion and CAH the air pockets were replaced by a lubricant oil that soaks into the textured substrate. This brings us to the second stage of fabrication.

**Stage II: Fabrication of biomimetic slippery liquid infused porous surface (SLIPS)**

In this stage, silicone oil is used as a lubricant fluid to produce a slippery surface from the fabricated superhydrophobic positive replica. Silicone oil is chosen because of its low volatile nature and its affinity with the textured surface which suppresses the volatile loss of the lubricant oil over the period of usage. To convert rough surface into a slippery surface, positive replica is immersed in silicone oil for some time (~4 hours) and then aligned vertically overnight over a support, to allow drainage of the excess lubricant under the action of gravity. Textured surface spontaneously soaks silicone oil onto the textured sites through capillary effect and replaces the air blanket with a layer a lubricant film. Further, sample is heated over a hot plate for 3 hours to evaporate any amount of excess lubricant from the substrate and this ensures that a thin and stable layer of lubricant is left behind over the substrate. Although, static contact angle, for the slippery surface, considerably reduces down to around 105° (similar to a normal PDMS) but at the same time, the CAH and SA values reaches to remarkably very low values (CAH ~ 2.0° and SA ~ 1.1°), confirming the super slippery nature of the surface (see Fig. 2b, Table S1, S2 and S3).

**Surface Characterization**

To characterize the fabricated surfaces, based on their surface topography, a field electron scanning electron microscope (FESEM) is used at 2 kV. The surface wettability characteristics were determined by measuring the contact angle hysteresis and sliding angles for DI water using an optical goniometer (Data-physics Instruments GmbH, Germany). Readings are taken at multiple locations as well as for various identical substrates and the average values are reported.

## 3. Results and discussion

In this article, thin micro/nano-textured rough surface was fabricated on PDMS substrate by replication of taro leaf surface structures. Textured surface was thereafter used to fabricate slippery liquid infused porous surfaces aiming at reduction in contact angle hysteresis and sliding angle values compared to both textured surface and the base material (PDMS). A brief comparison of the properties of different surfaces obtained during the fabrication is listed in Table 1.

**Table 1** Properties of different surfaces

|  | **Normal PDMS** | **Positive replica (Taro leaf)** | **Slippery surface** |
| --- | --- | --- | --- |
| **Surface wettability** | Hydrophobic | Superhydrophobic | Super-wetting |
| **Structure of material** | Irregular surface interface | Hierarchical and uniform scale roughness | Porous surface/gel |
| **Water CA (°)** | ~106 ± 1 | ~155 ± 1 | ~105.5 ± 1 |
| **CAH (°)** | ~7.0 - 8.0 | > 75 | ~2.1 ± 1 |
| **SA (°)** | > 10 | No sliding | ~1.1 ± 0.1 |

FESEM images show that negative replica of taro leaf comprises of uniformly distributed micro-pores over the surface, whereas the positive replica of taro plant leaf has micropillars present throughout the surface, and these micropillars are responsible for the superhydrophobic nature of positive replica (see Fig. 3a and 3b). When a liquid droplet is placed over this surface, air gets trapped, and act as a cushion for the overlying liquid droplet. This reduces the actual solid-liquid contact area which gives rise to superhydrophobicity even with a large CA. However, it has been observed that, this is also accompanied with an equivalently high CAH and SA caused by the pinning of liquid droplets over the textured surface. On Positive replica, at the top of micropillars, there exist nipple shaped mounds (*this is absent on the leaves of lotus and petals of rose, which exhibit extremely low and high CAH respectively*) which pierce through the boundary of the overlying liquid droplet. We envisage that this results in overall pinning of liquid droplets over the surface with very high adhesion. In Fig. 3c, FESEM images shows the topography of slippery surface fabricated from positive replica after infusing lubricant oil (*silicone*). We can observe that, slippery surface is having a flat, smooth and defect free interface which is responsible for low values of CAH and SA for SLIPS.

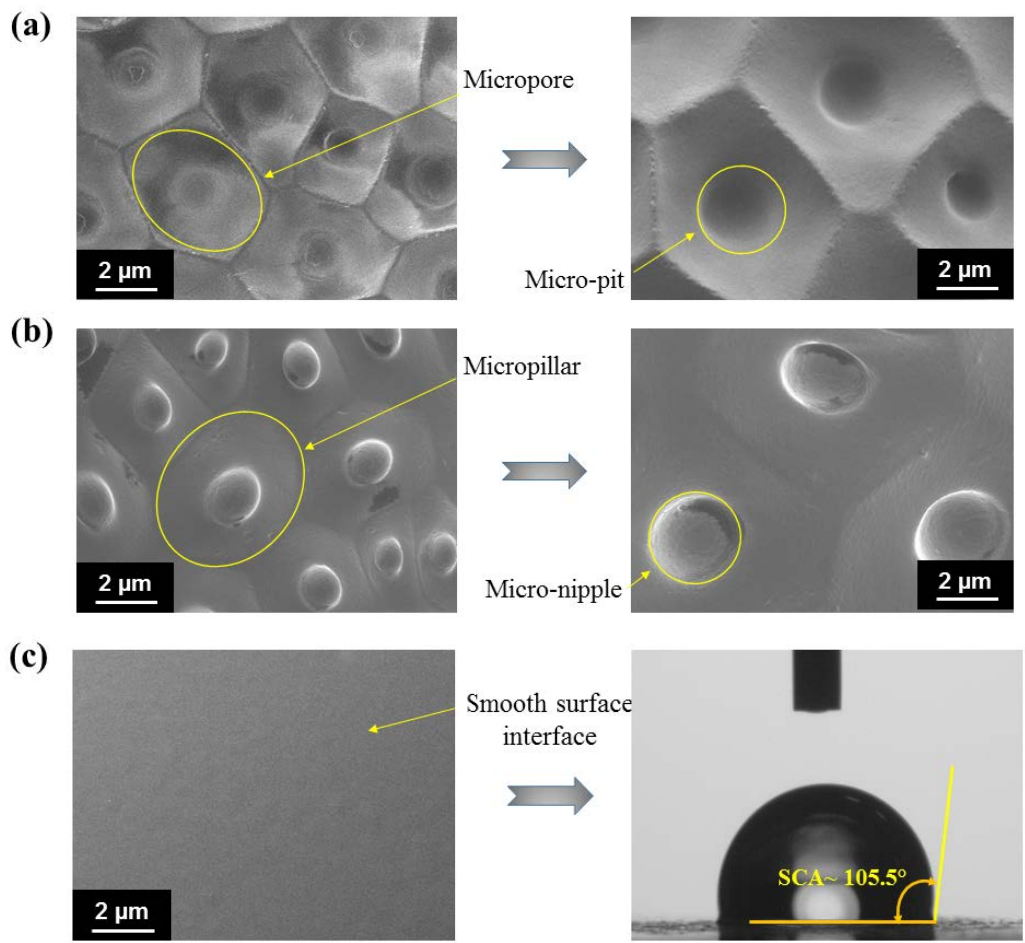

Fig. 3 FESEM images of (a) negative replica obtained from taro leaf template have micropore structure on the surface, (b) positive replica having micropillar structure with nipple kind of shape at the top each micropillar, (c) SLIPS have very smooth flat interface which is responsible for its super-slippery nature (although CA value for SLIPS is similar to CA of PDMS surface but the CAH value is extremely low ~2.0° whereas, PDMS surface have CAH ~8°).

**Stability of the slippery surface**

The stability as well as the robustness of the fabricated slippery surface is investigated via a three-step process which includes (a) shear stability - Excessive shear is known to result in abrupt loss of lubricant oils affecting its performance and therefore SLIPs are subjected to systematically varying shear. Other tests include (b) thermal stability and (c) droplet impact or mechanical stability.

*Shear stability test.* Shear test helped us in determining the robustness of the slippery surfaces under very high shear conditions. It is pertinent to determine the loss of lubricant for the particular shear rate and this is quantified by measuring the weight of the samples prior to the start of the experiment as well after the samples are sheared. To determine the shear stability of lubricant layer, the fabricated slippery substrates were subjected to continuous shear induced through spinning (using spin coater). The magnitude of the induced shear has been manipulated by controlling the rotational speed of the spin coater. In the present study, the substrates are exposed to spin rates 500 rpm to 3000 rpm in increments of 500 rpm for one minute each. Additionally, the surfaces are also characterized by contact angle goniometry where in the static equilibrium CAs, contact angle hysteresis and sliding angles are determined. These are correlated with the changes in the lubricant weight. It can be observed from Fig. 4 that there is no significant change in the sample weight (~0.945 %) (see ESI, Table S4) and CA, CAH and SA values also remained fairly constant. This implies that the wettability characteristics of the fabricated slippery surfaces are fairly stable and its slippery nature under the observed shear conditions is durable.

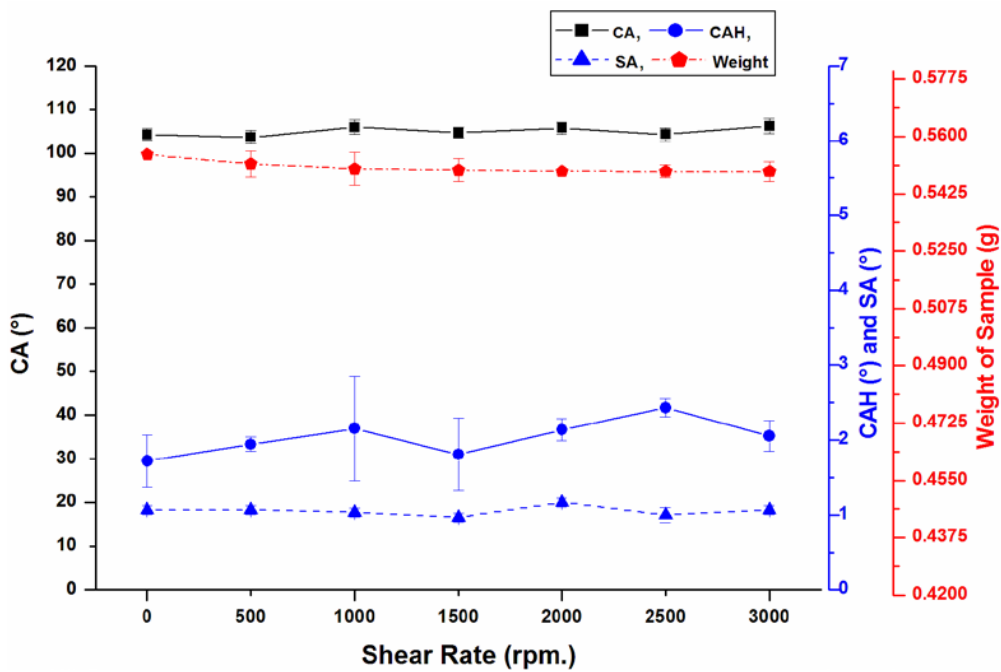

Fig. 4 Characterization of Substrate wettability (CA, CAH, SA) in the presence of the imposed shear rates. Correlation of the substrate stability with the loss of lubricant for the given shear rates. Fabricated surfaces are found to be relatively stable for the imposed shear rates.

*Thermal stability test*. Thermal stability herein implies the heat sustenance capacity of the lubricant oil infused into the pores of the textures at sufficiently high temperatures. It also includes maintenance of the surface properties, in particular surface hysteresis and contact angle at operating temperatures of ~ 90-95°C. It has been reported previously that the evaporation of lubricant oil from slippery surfaces generally affects their liquid-repellent and self-cleaning properties.[62] In order to assess the same for our samples, the fabricated surfaces were exposed to a (were placed in an oven) temperature of 95°C for a span of 7 days (a week). The sample weight was monitored, and its daily weight was measured. The weight loss in the lubricant oil, is outlined in Fig.5. The samples showed no apparent change (~0.267 %) in the weight after a week or during the week (Fig. 5). This can be attributed to extremely low volatility of the silicone oil (vapor pressure ~5 mmHg) and its affinity with the micro-textured surface which assists in binding it to the surface. It brings about a considerable decrease in the evaporative loss of the lubricant oil. Additionally, the measured CA, CAH and SA values were found to be unchanged over the period of evaporation. (~ minor variation, restricted within 2.3°). Thus, the surfaces are stable and functional even under elevated temperature conditions.

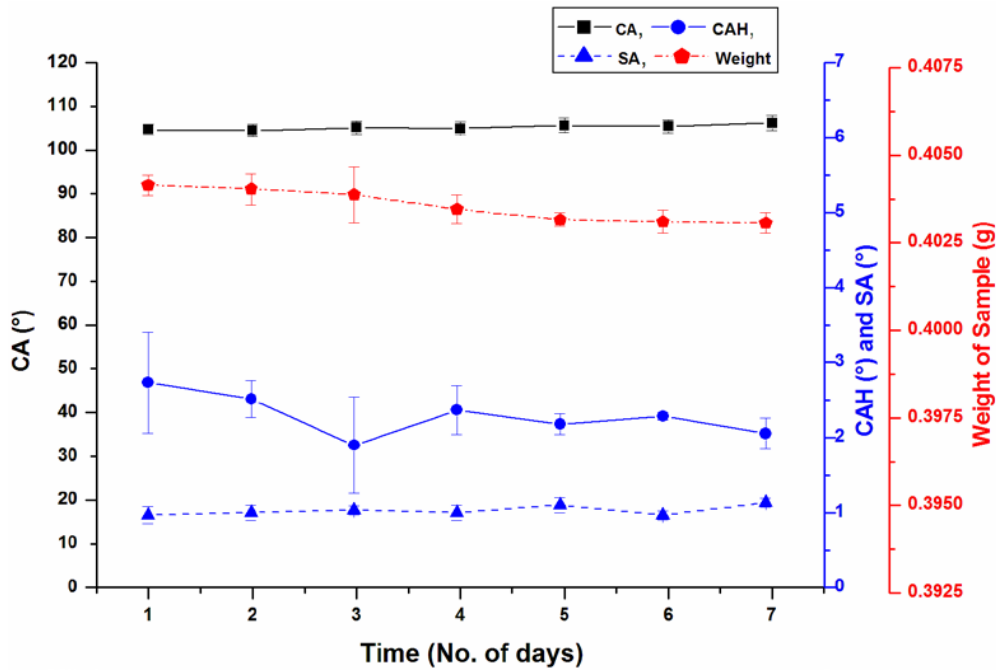

Fig. 5 Variation in the wettability characteristics from the slippery surface placed in a hot air oven at 95°C over the period of a week. The fabricated surface is thermally stable for the observed temperature range and time period.

*Droplet impact stability test.* Another major challenge in the usage of slippery surfaces lies in their stability against constant impact by the vertically bombardment of droplets, condensation of dew and fog in the outdoor environment[38]. Ideally, the impacting water layers must be repelled from the slippery surface without affecting the physical and chemical properties of the underlying lubricant film. To evaluate this, several water droplets were allowed to freely fall from a certain height (height was varied in increments of 1 - 8 cm) to create an impact on a particular location on the surface. This was repeated for multiple locations on the same substrate confirming no loss in stability in the presence of continuous impact. The surface wettability characteristics (CA, CAH and SA) and weight of the sample were recorded as a function of the impact height (1 - 8 cm) from the test surface. We observe that there is no appreciable change in the weight of the sample (~1.122 %) and the corresponding numerical values of CAH remain below 2.3° (Fig. 6), indicating excellent resistance of the fabricated surface against droplet impact (Table S6).

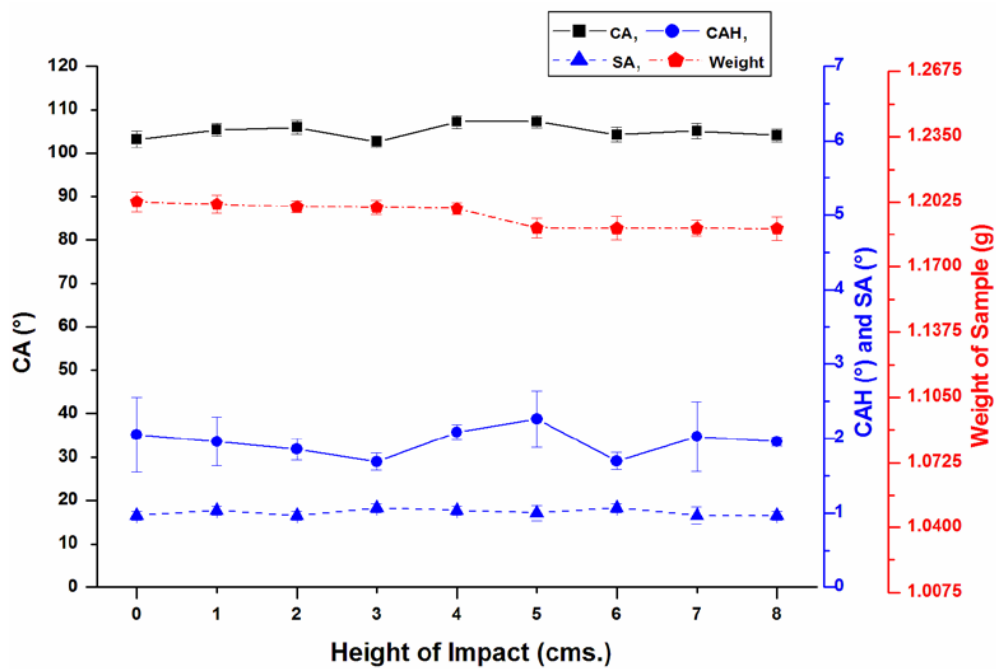

Fig. 6 Above graph shows the variations in CA, CAH, SA and sample weight for SLIPS against constant droplet impact (~ ten droplets on a particular location) on a particular spot with height of impact varying from 1 - 8 cm (location of droplet impact is different for each test height).

**Liquid-repellent ability of SLIPS**

The most remarkable ability of SLIPS is to repel a diverse range of liquids both organic and inorganic (water, dye, crude oil) as well as those of biological origin like blood. This makes it an ideal platform for microfluidic devices, MEMS and other commercial applications. To impart this property, it is essential that the surfaces possess low CAH apart from being flat, smooth and defect free at the liquid-solid interface to prevent pinning of the liquid droplets. This will result in the increase and hassle-free droplet mobility on SLIPS. The liquid droplets quickly slide-off from the slippery surface at sliding angles as low as ~1.2° (see, Fig. 7a). We attribute this to the presence of lubricant oil layer above the micro/nano-roughness sites of the slippery surface which prevents the pinning of three phase contact line by the liquid droplets. Also, the low values of CAH (~2.1°) ensure no pinning and high mobility of liquid droplets (see, Fig. 7b), irrespective of their low SCAs and large droplet-solid-surface contact area as compared to the textured surface[63]. Therefore, the slippery should possess a well-constructed micro/nanostructure network on the surface and also the lubricant oil should be completely immiscible with different liquids.

Large surface area of the micro/nano-textured surface enables the formation of a stable and thin film of lubricant oil, and these roughness sites are energetically favourable for the lubricant oil to spread into and to be retained by the textured surface for a longer duration of time. When liquid droplets were placed on a superhydrophobic positive replica (SUB) and SLIPS, then only SLIPS showed the liquid-repellent ability, while the liquid-droplets got pinned on a SUB surface despite having contact angle greater than 150°. In comparison to SLIPS, positive replica of taro leaf has larger adhesive forces because the motion of the liquid droplets did not extend the length of three phase contact line (TCL) and hence, results in the pining of droplets at the SUB surface. These liquid-repellent results demonstrate that slippery surfaces possess remarkable liquid-repellent ability, which enhanced the possibilities in variety of applications like in microchannels to reduce pressure drag, ship hulls to reduce frictional drag, deep-sea drilling rigs, biomedical devices, safer fuel transportation and anti-icing coatings by avoiding the coalesce of liquid-droplets at a place caused by the pining of tiny water droplets at the surface.

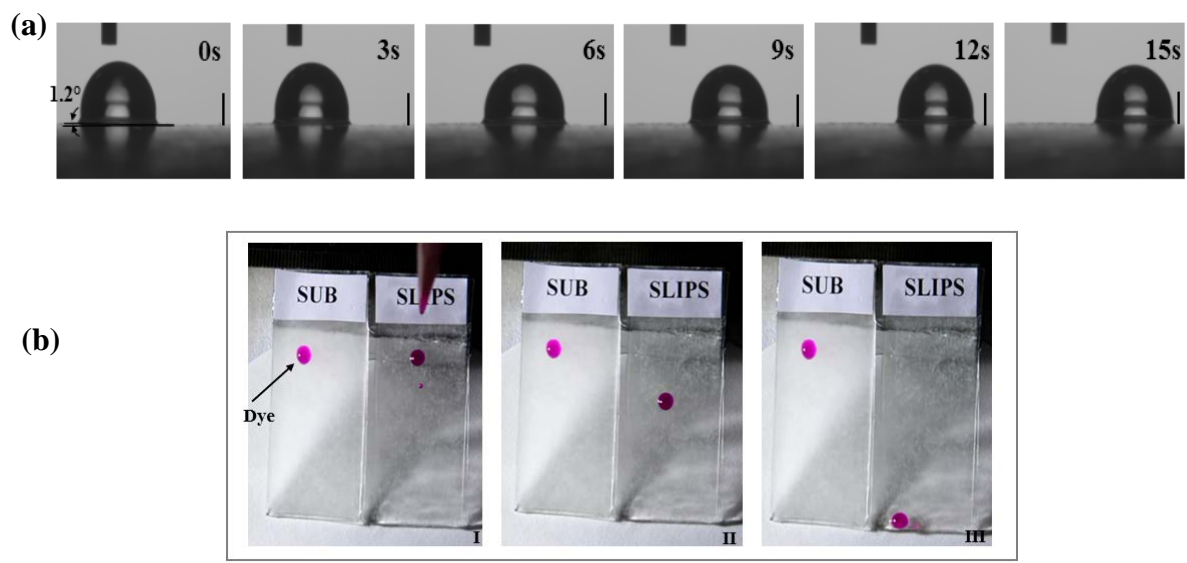

Fig. 7 (a) Dynamics of water droplet (~5 µL) movement on a SLIP surface at a very low sliding angle (~1.2°), (b) Representative demonstrations of liquid-repellent performance of the superhydrophobic positive replica of taro leaf (SUB) and the corresponding slippery surface (SLIPS) fabricated from it. (SLIPS repel water droplets efficiently as seen from the demonstration).

**Self-cleaning property of SLIPS**

It is essential that the SLIPs are capable of self-cleaning for them to be sustainable in varied working environments. To evaluate the same, we sprinkle graphite powder as sample dust on the

SLIPS followed by washings of de-ionized water droplets. These water droplets (see, Fig. 8). slide over the surface (comply to "slide to clean" strategy) and carry the dust particles along with them with considerable ease. Although, the sliding angles for water droplets for SLIPS are extremely low (~ 1.2°) it still functions effectively as a self-cleaning surface. The substrate material used for the fabrication of slippery surfaces is PDMS, and lubricant used is silicone oil which are known to have excellent optical clearness[42,64], making these, slippery surfaces potential candidates as a self-cleaning coatings in solar panels Thus, the fabricated surfaces in the present study are self-cleaning with respect to liquid and particulate matter.

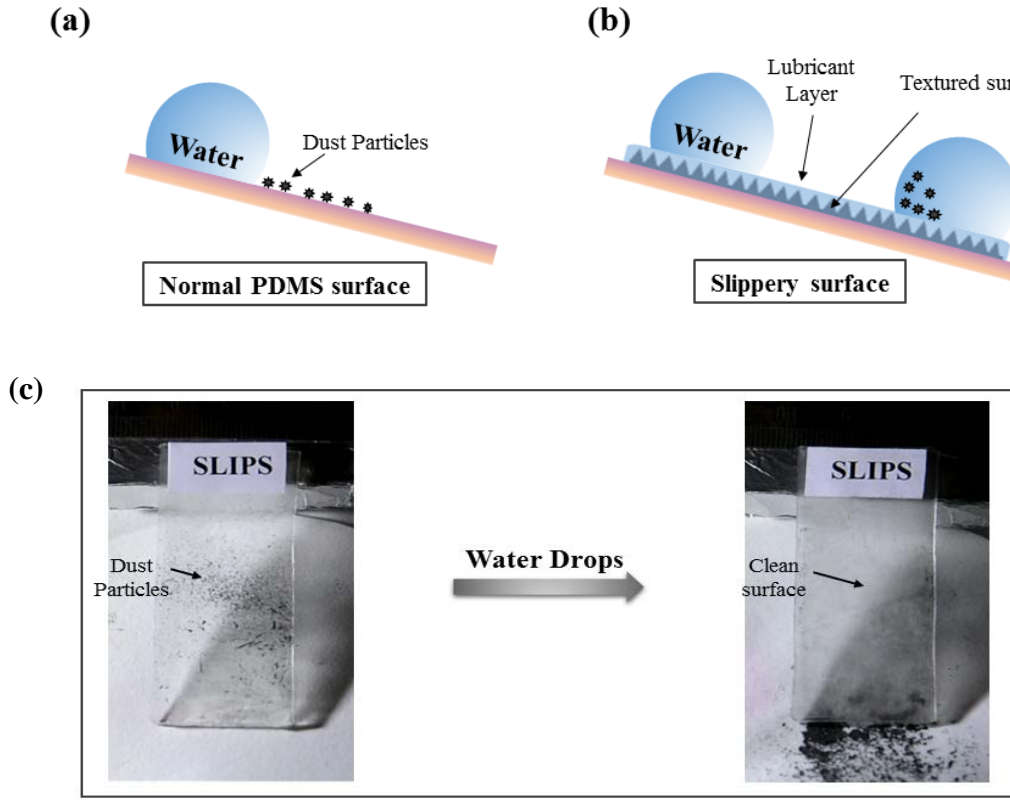

Fig. 8 Demonstration of self-cleaning ability of SLIPS, (a) Pinning of water droplets on a normal PDMS surface prevents free movement of a water droplet. (b) SLIPS - Water droplets slide freely with the dust particles along with it. "Slide-to-clean" cleaning mechanism exhibited by SLIPS, (c) Self-cleaning ability of SLIPS (*left*) was tested using dust (*graphite*) particles. Clean surface (*right*) shows impressive self-cleaning property exhibited by the slippery surface.

**Self-Repairing ability of SLIPS**

Besides liquid-repellent and self-cleaning properties, one of the most remarkable property of the SLIPS is their ability to self-repair within 0.1-1s after minor physical damage to the surface by abrasion or impact.[1,42].To evaluate self-repairing ability, we have tested SLIPS by manually applying number of sharp cuts on the surface using a knife. Sliding liquid droplets were observed before and after applying cuts, and we observed that there is no apparent change in the movement of liquid droplets. The numerical values of CAH < 2.3° and sliding angle remains below 1.1°. These results confirm the self-repairing ability of the slippery surfaces and we attribute the self-repairing behaviour of slippery surfaces to fluidic nature of the lubricant layer. Physical damage to a portion of surface creates a concentration gradient of lubricant oil, which along with surface-energy driven capillary forces enables the lubricant oil to simply flow towards the damaged portion and refill the physical voids formed by the damage. This process happens spontaneously imparting the self-repairing property to the SLIPs. (Fig. 9).

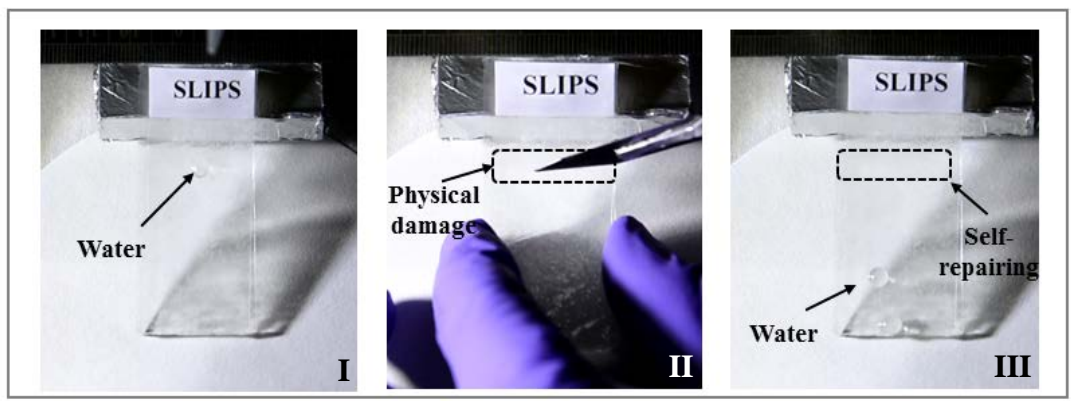

Fig. 9 Representative demonstrations of self-repairing property of SLIPS fabricated using taro leaf replica. Uninterrupted movement of a water droplet through the slippery surface even after periodic abrasion confirms the self-repairing ability of the SLIPS (Due to the fluidic nature of lubricant, capillary forces draws lubricant oil into the damaged area within very short time).

The fabricated SLIPS in the present study therefore possess the unique characteristics of extremely low values of contact angle hysteresis (< 2.2°) for water and very small sliding angles (~1.2°) along with self-cleaning, liquid-repellent, spontaneous self-repairing abilities and stability (shear, thermal and mechanical) in extreme working conditions which is hitherto unreported. In future, SLIPS can be developed to serve as a solution for meeting emerging demands in biomedical devices, fuel transportation, aerospace and marine applications, for self-cleaning purpose in solar-panels and windows, and in many other fields that are beyond the reach of current technologies.

## 4. Conclusions and outlook

In brief, this study presents an approach to fabricate a biomimetic slippery surface by infusing silicone oil into a textured surface which was prepared by obtained from taro leaf replica. The designed slippery surfaces exhibited stable and highly smooth lubricant layer with outstanding liquid-repellent and self-cleaning performance. Also, the fabricated slippery surface shows excellent self-repairing ability associated with the capillary action of lubricant oil presents in the micro-textured roughness sites. Furthermore, the synthesized slippery surface shows impressive long-term stability at extreme shear rates, against droplets impact on the surface and for temperature conditions as high as 95°C. As compared to the normal PDMS surface the values of CAHs (~ 2°) and SAs (~ 1.1°) for the fabricated SLIPS are extremely low which confirms their super-slippery nature. These slippery surfaces can be found useful in wide range of practical applications like slippery coatings of cars and buildings window glasses, solar-cell panels and sensors, bio-medical devices, micro-fluidic devices, marine and aerospace engineering among others.

The future scope is incorporating various cutting-edge features like anti-reflection, bio-logical stability, high-pressure stability, pest-repellent, enhanced oil recovery and non-toxic slippery coatings for kitchenware, etc. into slippery surfaces. From the engineering point of view, future challenges are towards commercialization of SLIPS technology by development of surfaces which are more cost-effective, flexible in use, reliable, durable and sustainable to increase their wide utility.

## Acknowledgements

The authors would like to acknowledge Sponsored Research and Industrial Consultancy (SRIC) Cell, IIT Kharagpur for the financial support of the 'Plant on-a-Chip' project provided through the SGDRI grant.

## Notes and references

# Electronic Supplementary Information

# Pitcher Plant Inspired Biomimetic Liquid Infused Slippery Surface Using Taro Leaf

Rahul Sharma,[a†] Sankha Shuvra Das,[a†] Udita Uday Ghosh,[b] Sunando DasGupta,[b] and Suman Chakraborty[a]*,

[a]Department of Mechanical Engineering, Indian Institute of Technology Kharagpur, Kharagpur-721302, India.
[b]Department of Chemical Engineering, Indian Institute of Technology Kharagpur, Kharagpur-721302, India.

[‡]Authors have equally contributed to this work.

*email: suman@mech.iitkgp.ernet.in

(a) 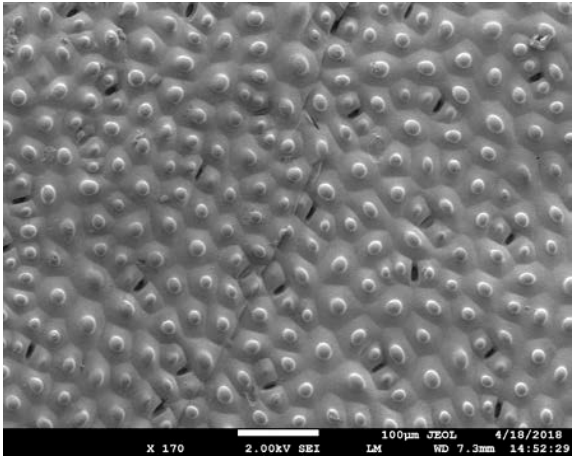

(b) 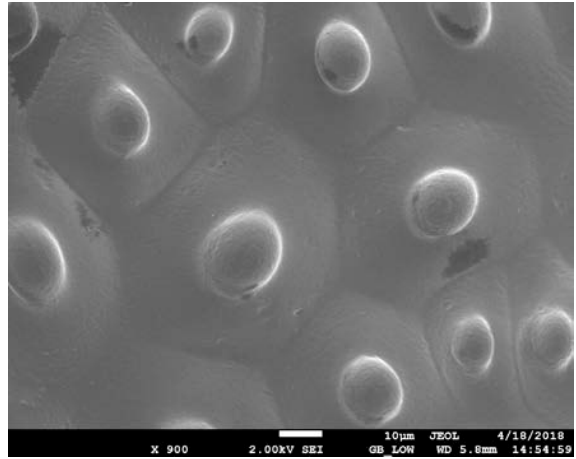

(c) 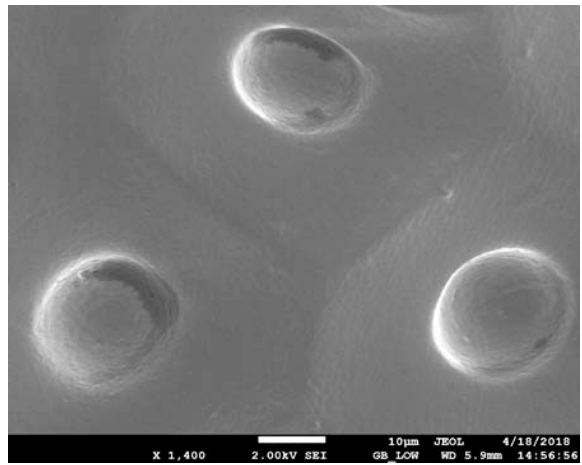

**Figure S1**. FESEM images of superhydrophobic positive replica of taro leaf shows surface microstructures at different magnifications, **(a)** X 170, **(b)** X 900, **(c)** X 1,400

**Table S1.** Different characterization parameters of a slippery surface measured after placing the sample in a hot air oven at 95°C for a week to analyse the changes occurred due to evaporation of lubricant oil.

| Time (Day) | CA (°) | St. Deviation (CA) | CAH (°) | St. Deviation (CAH) | SA (°) | St. Deviation (SA) |
|---|---|---|---|---|---|---|
| 1 | 104.757 | 1.21433 | 2.735 | 0.67332 | 0.966 | 0.11547 |
| 2 | 104.472 | 1.42657 | 2.517 | 0.24564 | 1.000 | 0.10000 |
| 3 | 105.158 | 1.53713 | 1.900 | 0.64352 | 1.033 | 0.05774 |
| 4 | 104.976 | 1.61928 | 2.367 | 0.32405 | 1.000 | 0.10000 |
| 5 | 105.723 | 1.62145 | 2.181 | 0.13623 | 1.100 | 0.10000 |
| 6 | 105.496 | 1.51689 | 2.285 | 0.05225 | 0.966 | 0.05774 |
| 7 | 106.255 | 1.74482 | 2.054 | 0.20438 | 1.133 | 0.05774 |

**Table S2.** Characterization parameters of the sample measured after droplets impact with height of impact varying from 1-8 cm.

| Height (cm.) | CA (°) | St. Deviation (CA) | CAH (°) | St. Deviation (CAH) | SA (°) | St. Deviation (SA) |
|---|---|---|---|---|---|---|
| 0 | 103.200 | 1.86153 | 2.053 | 0.49929 | 0.966 | 0.05774 |
| 1 | 105.332 | 1.42692 | 1.961 | 0.32106 | 1.033 | 0.05774 |
| 2 | 105.932 | 1.61898 | 1.860 | 0.13872 | 0.966 | 0.05774 |
| 3 | 102.610 | 1.39733 | 1.694 | 0.11845 | 1.067 | 0.05774 |
| 4 | 107.183 | 1.51726 | 2.085 | 0.1007 | 1.033 | 0.05774 |
| 5 | 107.180 | 1.44494 | 2.261 | 0.37356 | 1.000 | 0.10000 |
| 6 | 104.278 | 1.64999 | 1.701 | 0.11769 | 1.067 | 0.05774 |
| 7 | 105.097 | 1.72515 | 2.027 | 0.46596 | 0.967 | 0.11547 |
| 8 | 104.142 | 1.43784 | 1.960 | 0.05551 | 0.967 | 0.05774 |

**Table S3.** Different Characterization parameters of the sample measured after spinning the sample from 500-3000 rpm. for 1min. at each spin rate.

| Shear (rpm.) | CA (°) | St. Deviation (CA) | CAH (°) | St. Deviation (CAH) | SA (°) | St. Deviation (SA) |
|---|---|---|---|---|---|---|
| 0 | 104.251 | 1.37755 | 1.722 | 0.34740 | 1.067 | 0.05774 |
| 500 | 103.686 | 1.51025 | 1.944 | 0.09602 | 1.067 | 0.05774 |
| 1000 | 106.014 | 1.61728 | 2.155 | 0.69508 | 1.033 | 0.05774 |
| 1500 | 104.697 | 1.19945 | 1.809 | 0.48073 | 0.967 | 0.05774 |
| 2000 | 105.751 | 1.37138 | 2.139 | 0.14673 | 1.167 | 0.05774 |
| 2500 | 104.305 | 1.48345 | 2.433 | 0.11980 | 1.000 | 0.10000 |
| 3000 | 106.255 | 1.67448 | 2.054 | 0.20438 | 1.067 | 0.05774 |

**Table S4.** Variation in the weight of the slippery samples with shear rate shows no apparent loss of lubricant oil even at extreme shear conditions

| Shear Rate (rpm.) | Weight of Sample (g) | Std. Deviation |
|---|---|---|
| 0 | 0.55457 | 5.77E-5 |
| 500 | 0.55170 | 4.12E-3 |
| 1000 | 0.55020 | 5.13E-4 |
| 1500 | 0.54982 | 3.52E-4 |
| 2000 | 0.54947 | 2.39E-4 |
| 2500 | 0.54942 | 5.42E-3 |
| 3000 | 0.53933 | 3.47E-4 |

**Table S5.** Very minimal changes in the weight of the sample placed in an oven at 95°C constantly for a week shows the thermal stability of the slippery surface.

| Time (No. of days) | Weight of Sample (g) | Std. Deviation |
|---|---|---|
| 1 | 0.40415 | 3.12E-4 |
| 2 | 0.40404 | 4.53E-4 |
| 3 | 0.40388 | 8.29E-4 |
| 4 | 0.40347 | 4.03E-4 |
| 5 | 0.40317 | 2.14E-4 |
| 6 | 0.40311 | 3.39E-4 |
| 7 | 0.40307 | 3.01E-4 |

**Table S6.** Consistency in the weight of the sample shows the stability of the fabricated SLIPS against droplets impact.

| Height of impact (cm.) | Weight of Sample (g) | Std. Deviation |
|---|---|---|
| 0 | 1.20257 | 5.07E-3 |
| 1 | 1.20123 | 4.57E-3 |
| 2 | 1.20007 | 1.03E-3 |
| 3 | 1.19977 | 3.53E-3 |
| 4 | 1.19923 | 3.74E-3 |
| 5 | 1.18953 | 5. 13E-3 |
| 6 | 1.18933 | 6.01E-3 |
| 7 | 1.18929 | 4.01E-3 |
| 8 | 1.18908 | 6.19E-3 |